# Temperature-Insensitive Analog Vector-by-Matrix Multiplier Based on 55 nm NOR Flash Memory Cells

X. Guo[1], F. Merrikh Bayat[1], M. Prezioso[1], Y. Chen[2], B. Nguyen[2], N. Do[2], and D. B. Strukov[1]

[1] UC Santa Barbara, Santa Barbara, CA 93106-9560, U.S.A.
[2] Silicon Storage Technology Inc., A Subsidiary of Microchip Technology Inc., San Jose, CA 95134, U.S.A

*Abstract*—We have fabricated and successfully tested an analog vector-by-matrix multiplier, based on redesigned 10×12 arrays of 55 nm commercial NOR flash memory cells. The modified arrays enable high-precision individual analog tuning of each cell, with sub-1% accuracy, while keeping the highly optimized cells, with their long-term state retention, intact. The array has an area of 0.33 µm² per cell, and is at least one order of magnitude more dense than the reported prior implementations of nonvolatile analog memories. The demonstrated vector-by-vector multiplier, using gate coupling to additional periphery cells, has ~2% precision, limited by the aggregate effect of cell noise, retention, mismatch, process variations, tuning precision, and capacitive crosstalk. A differential version of the multiplier has allowed us to demonstrate sub-3% temperature drift of the output signal in the range between 25 °C and 85 °C.

*Keywords—Analog computing, NOR flash memory, temperature drift, vector-by-matrix multiplier*

## I. Introduction

Analog circuits are very appealing for signal processing at relatively low precision requirements, for example artificial neural networks, because they may far overcome digital circuits of the same functionality in circuit density, speed and energy efficiency [1, 2]. However, prior designs of individually adjustable memory cells for such analog circuits, for example "synaptic transistors" [3], which are based on standard CMOS process, had relatively large cells (of area ~$10^3\ F^2$, where $F$ is the minimum feature size), leading to larger time delays and power consumption [4-6].

An alternative way forward has been enabled by the progress of the industrial flash memory technology, now featuring highly optimized floating-gate cells, which may be embedded into CMOS integrated circuits. For example, Fig.1 shows the "supercell" of the advanced commercial 55 nm ESF3 NOR flash memory from SST Inc. [7]. However, the original ESF3 technology is not suitable for analog applications, because its arrays do not allow for precise individual tuning of the state (floating gate charge) of each cell. In this paper, we report a redesign of the ESF3 cell arrays, which enables such tuning. The redesigned arrays were used to demonstrate a small vector-by-matrix multiplier operating in the low-power subthreshold mode, with gate coupling of array cells to the input (peripheral) cells. In order to reduce the temperature drift, pertinent to the subthreshold operation of the cells, we have implemented and characterized a differential version of the multiplier, which minimizes the output signal drift.

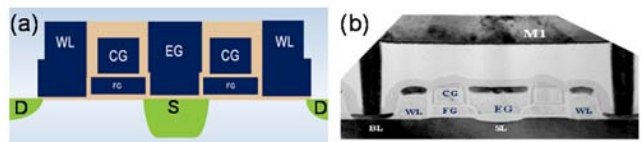

**Fig. 1.** SST's 55 nm ESF3 NOR flash memory cells: (a) schematic view, and (b) TEM image of the cross-section of a "supercell" incorporating two floating-gate transistors with a common source (S) and erase gate (EG) [7].

## II. Original and Modified Memory Cell Arrays

### A. Original Array and Cell Programming

The SST NOR flash memory is based on "supercells" with two floating-gate transistors sharing the source (S) and the erase gate (EG), but are controlled by different word-line (WL) and coupling (CG) gates - see Fig. 1. In the original ESF3 memory arrays, the cells are connected as Fig. 2a shows, with six row lines per supercell, connecting transistor sources, erase gates, coupling gates, and word-line gates, while each column has only one ("bit") line connecting transistor drains (D).

In this array topology, each cell may be programmed individually, by hot-electron injection into its floating gate. For that, the voltage on the source line (SL in Fig. 2) of the cell's row is increased to 4.5 V (while those in other rows are kept at 0.5 V), with the proper column selected by lowering the bit line (BL) voltage to 0.5 V (while keeping all other bit line voltages above 2.25 V). This process works well for providing proper digital state, with 1- or even 2-bit accuracy. However, it is insufficient for cell tuning with analog (say, 1%) precision. Unfortunately, the reverse process ("erasure"), using the Fowler-Nordheim tunneling of electron from the floating gates to the erase gates, may be performed, in the original arrays, only in the whole row, selected by applying a high voltage of ~ 11.5

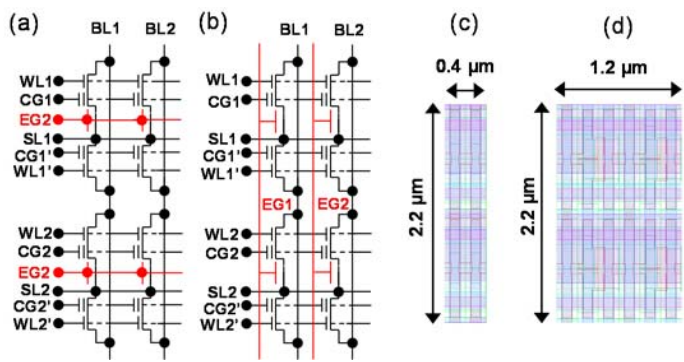

**Fig. 2.** (a, c) Original and (b, d) modified circuitry and layout for 4-supercell ESF3 55-nm memory array.

V to the corresponding erase gate line (with all other EG voltages kept at 0 V). So, these arrays do not allow for a precise analog cell tuning, which unavoidably requires an iterative, back-and-forth (program-read-erase-read-program...) process, with the run-time result control.

*B. Array Modification and Cell Tuning*

We have modified the ESF3 memory arrays as shown in Fig. 2b, by connecting the erase gates of all cells of one column with an additional line, while eliminating the row lines connecting these gates. (Note that this redesign is different from the one performed by our group earlier [8] with the 180 nm ESF1 technology, because of a different structure of its supercells.)

In the modified arrays, the analog hot-electron programming of each cell may be performed by applying 10 μs pulses of a fixed amplitude of 4.5 V to the source line of the corresponding row. In this process, the proper column is selected by applying a positive voltage ~4 V between the erase-gate and bit lines, while keeping this voltage negative for all un-selected columns [10]. Fig. 3a documents the inhibition of the unwanted programming process in a half-selected cell at the increase of the bit-line (i.e. drain) voltage.

The opposite process of individual analog erasure via the Fowler-Nordheim tunneling is now also possible, by using the new column lines to apply high-amplitude (11.5 V), 0.5 ms pulses to the erasure gates of the selected column. The proper row is selected by grounding the corresponding coupling gate line, while keeping a high voltage (+8 V) on these lines of unselected rows. As Fig. 3b shows, such a positive bias inhibits the Fowler-Nordheim tunneling in half-selected cells, due to a relatively high capacitance between the coupling gate and the floating gate of the same transistor [11].

Due to the line rerouting, the array area per cell has nearly tripled – cf. Figs. 2c and 2d. However, even with this increase the area is still as small as 0.33 μm$^2$, i.e. ~ 110 $F^2$, much smaller than in any other design we are aware of.

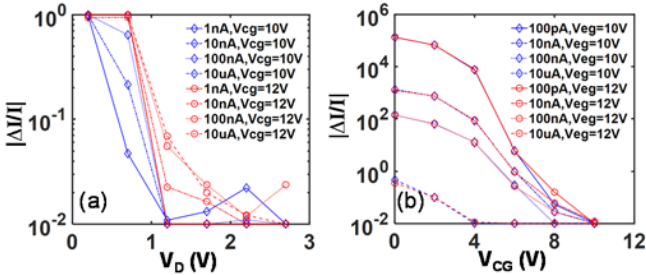

**Fig. 3.** (a) Programming inhibition and (b) erasure inhibition in the transistors of half-selected cells. Unless specified otherwise, the shown readout (source-to-drain) currents have been measured at $V_{WL}$ = 2.5 V, $V_{CG}$ = 2.5 V, $V_D$ = 1 V, $V_S$ = 0 V, and $V_{EG}$ = 0 V.

*C. Subthreshold Operation*

The ability to tune the floating gate cells of the modified arrays continuously is illustrated in Fig. 4 which shows the readout current as a function of the coupling gate voltage for a selected equidistant series of cell states. These semi-log plots have wide quasi-linear segments, corresponding to the nearly-exponential behavior of the current in the subthreshold region. In the current range from 100 pA and 30 nA, the subthreshold slope factor $n$, defined by the well-known relation $I \propto \exp\{qV_{CG}/nk_BT\}$, varies only from 5 to 5.1 for all the 15 states shown in Fig. 4. This low variability of $n$ enables the implementation of highly linear signal transfer in gate-coupled current mirrors using these cells [6].

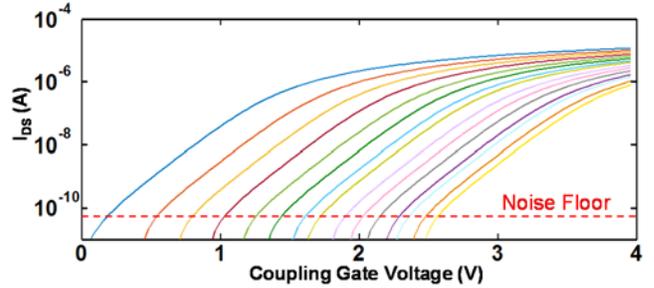

**Fig. 4.** Drain-source current of a modified ESF3 cell as a function of the coupling gate voltage, for several different memory states.

*D. Analog State Retention and Noise*

The ESF3 flash technology guarantees a 10-year digital-mode retention at temperatures up to 125˚C [7]. To explore its analog mode retention, we have programmed 7 memory cells to 7 different states from around 100 pA to 100 nA covering the whole subthreshold region, and then were continuously monitoring their output current within a day under 85 ˚C as shown in Fig. 5a. Each point on this panel is an average over 128 samples taken during 16 ms periods. Fig. 5b shows the relative r.m.s. variation of the current during the measurement period for the 7 states shown in Fig. 5a. For larger currents the variation is below 1%, increasing to ~4% only at the lower boundary of the range.

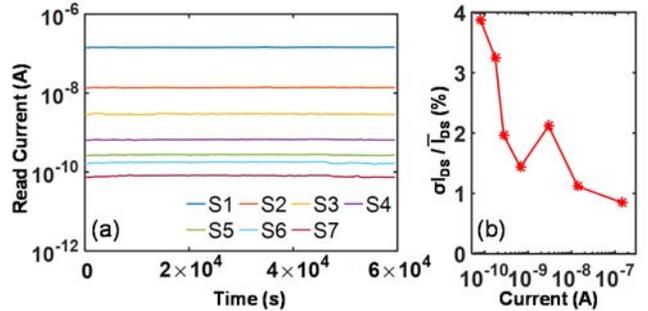

**Fig. 5.** (a) Retention measurements for several cells in different memory states at 85°C, and (b) the average relative variation of the readout currents during this time interval.

*E. Temperature Dependence of the Readout Current*

In order to fairly characterize the temperature dependence of the cell output current in the subthreshold region, we have programmed 8 cells to 8 different states, equally spread over the useful dynamic range. Then, in 3 different experiments, appropriate coupling gate voltages were applied to each cell, to make the readout currents of them all equal to, sequentially, 1 nA, 10 nA and 100 nA at 25 ˚C. After that, temperature was ramped up from 25˚C all the way to 85˚C, and the readout current of each cell was monitored. Fig. 6 shows the results of these 3 experiments. In accordance with our expectations (and the measured values of $n$), the currents increased significantly – more than by an order of magnitude for the lowest initial current.

This work was supported by DARPA's UPSIDE program under contract HR0011-13-C-0051UPSIDE via BAE Systems.

Though in the gate-coupling scheme (see below) this changes are mostly compensated by similar changes in the input (peripheral) transistors, this fact still shows that the temperature sensitivity of the subthreshold current requires special attention – see Sec. IV below.

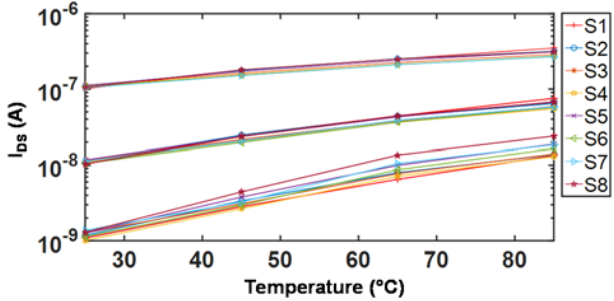

**Fig. 6.** Temperature dependence of drain-source current for several memory cells with different memory states, with the initial readout currents close to 1 nA, 10 nA and 100 nA.

### III. VECTOR-BY-MATRIX MULTIPLIER

To implement the vector-by-matrix multiplication, we have used the gate coupling of the tunable floating gate cells of each row of the array with a similar "peripheral" cell, with the virtual-bias condition imposed (by external circuitry) on the output (column) wires [1, 6] (Fig. 7).

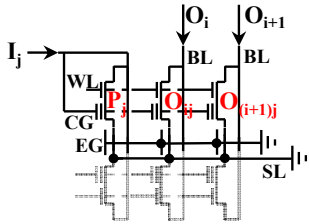

**Fig. 7.** The vector-by-matrix multiplication scheme based on gate coupling of the floating-gate cells. (For clarity, only one peripheral (P) and two array (O) cells of the same ($j^{th}$) row are shown.

Since all cells of the same row share the same coupling gate voltage $V_j$, in the subthreshold operation mode the $j$-th component $O_{ij}$ of the output current $O_i$ in the $i$-th column is proportional to the input current $I_j$ in the $j$-th-row:

$$I_j = I_0 \exp\left\{q\frac{V_j - V_{th}^{(j)}}{nk_BT}\right\},$$

$$O_i = \sum_j O_{ij} = \sum_j I_0 \exp\left\{q\frac{V_j - V_{th}^{(i,j)}}{nk_BT}\right\} \equiv \sum_j w_{ij} I_j, \quad (1)$$

with current-independent proportionality coefficients $w_{ij}$, which are determined by the differences of threshold voltages $V_{th}$ of the array cells and the peripheral transistors:

$$w_{ij} = \exp\left\{q\frac{V_{th}^{(j)} - V_{th}^{(i,j)}}{nk_BT}\right\}. \quad (2)$$

In turn, each threshold voltage is determined by the analog state (physically, the floating gate charge) of the cell, so that each $w_{ij}$ may be adjusted to the desirable value (typically, below 1).

Thus the fundamental Kirchhoff's current law enables the implementation of the analog vector-by-matrix multiplication [4, 8]. Fig. 8 shows our test 10×10 multiplier based on the modified ESF3 cell array. The peripheral cells are located in the additional columns on the left and the right from the basic array. (Two columns are necessary because with the ESF3 supercell structure we can use only one half of each supercell as the peripheral cell.)

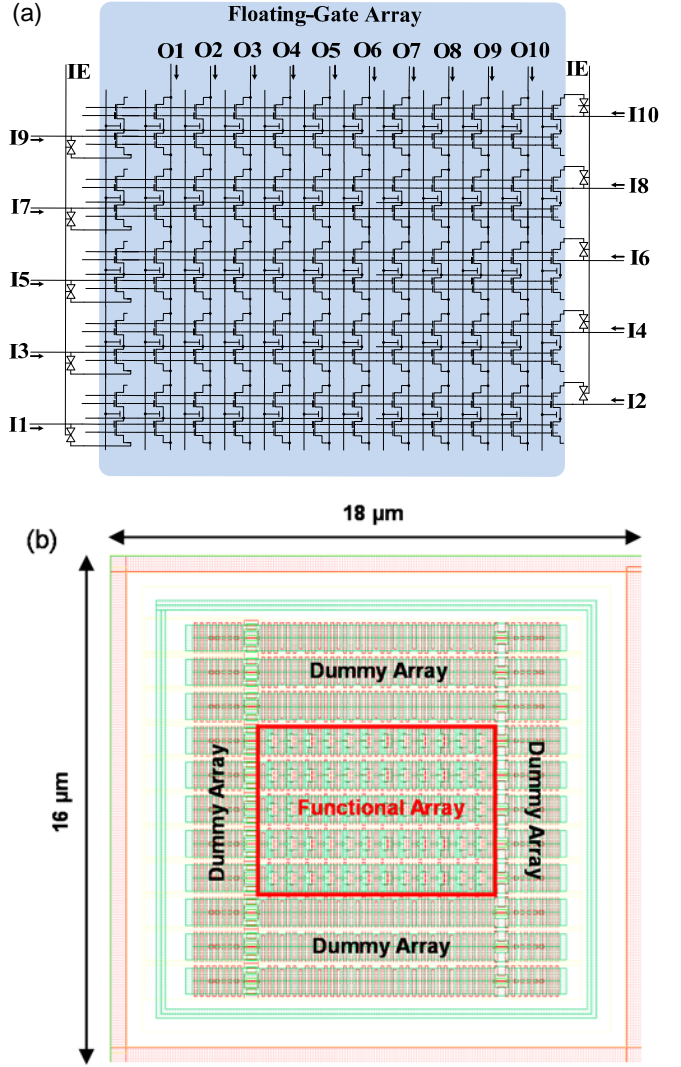

**Fig. 8.** Gate-coupled vector-by-matrix multiplier based on a 10×(10+2) array of ESF3 floating-gate cells, together with auxiliary pass-gates (which are disabled during tuning with IE signal): (a) schematics; (b) layout for 55-nm fabrication.

Fig. 9 illustrates the analog tuning capability of the array. All 10×10 array cells have been tuned one-by-one by an automatic feedback controlled application of alternating programming pulses to their source electrodes and erasing pulses to their erase gates. After each tuning pulse, the external control circuitry read out the cell output current at standard bias conditions, and made a decision about the next pulse's destination and amplitude, until the read-out current has reached the target value with the 5% precision [9]. Fig. 9a shows the results of 3 separate experiments of tuning all 100 cells of the array to different target values of the output current: 1 nA (red line), 100 nA (green line), and an exponential function of the cell number, within the rage from 100 pA to 1 μA (yellow line). Fig. 9b shows the relative errors achieved in last experiment. The data mean that larger tuning errors (of the order of 10%) take place for smaller target currents, because of the relative large intrinsic noise of the devices.

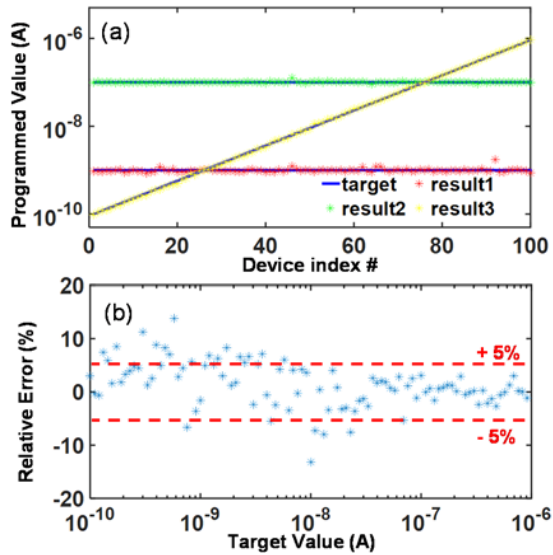

**Fig. 9.** (a) Measured versus target weights for 100 devices, and (b) Measured tuning error for 100 devices at a tuning precision target of 5%.

As a simple illustration of multiplier's operation, Fig. 10 shows the results of multiplication of 4 input signals by 4 different weights: $w_1 = 0.25$, $w_2 = 1$, $w_3 = 0.5$, and $w_4 = 0.125$, performed by 4 cells of one column of the array, tuned with a 1% precision. This experiment demonstrates that the relative error, incorporating contributions from all sources (device noise, state retention, impedance mismatch, parameter variation, tuning precision, and capacitive crosstalk) does not exceed 2%.

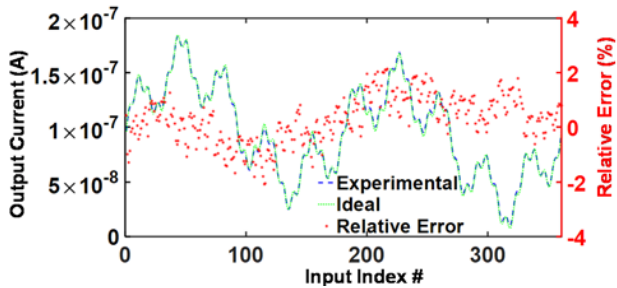

**Fig. 10.** Ideal (green line) and real (blue dashes) outputs at a 4-input vector-by-vector multiplication, and their difference (red dots). The four inputs are quasi-DC currents sampled from sine function 50 nA × [1+ Sin(2π × Input Index# × $f$)], with $f = 1/8$, 1/36, 1/180, and 1/360.

## IV. TEMPERATURE INSENSITIVE OPERATION

According to Eq. (2), in the coupled-gate operation mode, much of the thermal dependence of the subthreshold current is compensated, but besides the special case $w_{ij} = 1$, the compensation is incomplete. Indeed, our measurements have confirmed that in agreement with this relation, that as temperature is raised from 25°C to 85°C, weight $w_{ij}$, initially equal to 0.9, increases by ~10%.

However, there is a straightforward way to decrease the temperature sensitivity, at the cost of a two-fold increase of hardware. For that, one can subtract output currents of two cells (say, those shown in Fig. 7), with their individual weights tuned to, respectively, ($w_b + w/2$) and ($w_b - w/2$). Here $w$ is the desired net weight, and $w_b$ is the "bias weight", which may be optimized to suppress the temperature dependence of the new output current. A straightforward analysis of this scheme, using Eq. (2), shows that after such optimization, the temperature drift of the output may be reduced to less than 1% at the [25°C, 85°C] interval, for any weight $0 < w_{ij} < 1$. Fig. 11 shows the results of our preliminary experiments with this mode, showing the drifts not exceeding 2.7% in that temperature interval.

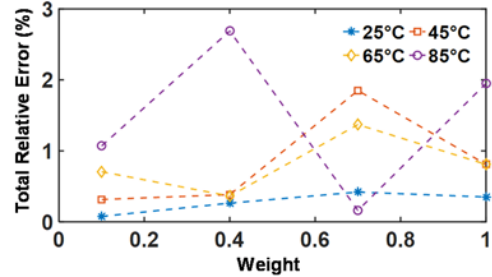

**Fig. 11.** The total relative error of the reproduction of a 100 nA input signal for several values of $w$, at various temperatures.

## V. SUMMARY AND DISCUSSION

We have demonstrated a simple prototype of an extremely compact analog vector-by-matrix multiplier, with a-few-percent precision and temperature drift, based on redesigned arrays of the commercial ESF3 NOR flash memory. While we have not yet directly measured the multiplier's bandwidth and energy efficiency, because of our current experimental setup limitations, our estimates, based on experimentally measured parameters of the cells, show that these figures-of-merit should be several orders of magnitude better than those of state-of-the-art digital multipliers with similar precision.

## ACKNOWLEDGMENTS

Useful discussions with P.-A. Auroux, J. Edwards, M. Graziano, and K. Likharev are gratefully acknowledged.